\documentclass[conference]{IEEEtran}
\usepackage{xcolor}
\usepackage{listings}
\usepackage{multicol}
\usepackage[ruled]{algorithm2e}

\ifCLASSINFOpdf
  \usepackage[pdftex]{graphicx}
\else
\fi

\usepackage{amsmath}
\usepackage{graphicx}

\usepackage{xspace}
\usepackage{multirow}
\usepackage{mwe}
\usepackage{hyperref}

\usepackage{amssymb}  
\usepackage{pifont}
\usepackage{threeparttable}

\usepackage{balance}

\usepackage{xcolor}
 \definecolor{shadecolor}{rgb}{0.98,0.98,0.38}  
\usepackage{framed} 
\usepackage{soul}

\newcommand{\myparagraph}[1]{\textbf{\textbf{#1 }}}

\usepackage{textcomp}

\usepackage{xcolor}

\soulregister\Hl{7}
\soulregister\ref7
\soulregister\cite7
\soulregister\pageref7
\soulregister\sysname7
\soulregister\algname7
\soulregister\myparagraph7
\soulregister\Hl{7}
\soulregister\caption7

\begin{document}
\newcommand{\sysname}{SPAC\xspace}
\title{\sysname: 
Automating FPGA-based Network Switches with Protocol Adaptive Customization
}

\author{
\IEEEauthorblockN{Guoyu Li$\dagger$, Yang Cao$\dagger$, Lucas H L Ng$\dagger$, Alexander Charlton$\dagger$, Qianzhou Wang$\dagger$, Will Punter$\dagger$, \\Philippos Papaphilippou$\ddagger$, Ce Guo$\dagger$, Hongxiang Fan$\dagger$, Wayne Luk$\dagger$, Saman Amarasinghe$\S$ and Ajay Brahmakshatriya$\S$}
\IEEEauthorblockN{$\dagger$ Imperial College London, $\ddagger$ University of Southampton, $\S$ Massachusetts Institute of Technology}
\IEEEauthorblockA{\{g.li25, yang.cao24, lucas.ng22, alexander.charlton22, qianzhou.wang17, will.punter23, c.guo,\\ hongxiang.fan, w.luk\}@imperial.ac.uk; P.Philippos@soton.ac.uk; saman@csail.mit.edu; ajaybr@mit.edu}
}

\maketitle
\begin{abstract}

\looseness=-1
With network requirements diverging across emerging applications, latency-critical services demand minimal logic delay, while hyperscale training and collectives require sustained line-rate throughput for synchronized bulk transfers. 
This divergence creates an urgent need for custom network switches tailored to specialized protocols and application-specific traffic patterns.
This paper presents \sysname (Switch and Protocol Adaptive Customization), a novel approach that automates the generation of FPGA-based network switches co-optimized for custom protocols and application-specific traffic patterns.
\sysname introduces a unified workflow with a domain-specific language (DSL) for protocol-architecture co-design, a library of modular HLS-based adaptive switch components, and a trace-aware Design Space Exploration (DSE) engine.
By providing a multi-fidelity simulation stack, \sysname enables rapid identification of Pareto-optimal designs prior to deployment. 
We demonstrate the efficacy of the domain-specific adaptation of \sysname across a spectrum of real-world scenarios, spanning from latency-sensitive sensor and HFT networks to hyperscale datacenter fabrics. Experimental results show that by tailoring the micro-architecture and protocol to the specific workload, \sysname-generated designs reduce LUT and BRAM usage by 55\% and 53\%, respectively. Compared to fixed-architecture counterparts, \sysname delivers latency reductions ranging from 7.8\% to 38.4\% across various tasks while maintaining adequate resource consumption and packet drop rate. \footnote{Our design is open source at: \href{https://github.com/spac-proj/SPAC}{https://github.com/spac-proj/SPAC}} 

\end{abstract}

\section{Introduction}

\begin{figure}[tbp] 
\centering
\includegraphics[width=0.9\linewidth]{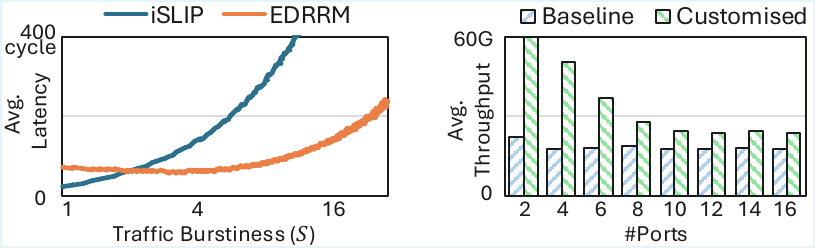} 
\vspace*{-2ex} 
\caption{Hardware Sensitivity (left): different scheduler architectures favor different traffic patterns. Protocol Sensitivity (right): throughput comparison of a \sysname switch using a standard protocol versus a custom protocol.} \label{fig:motivation}
\vspace*{-4ex} 
\end{figure}

Modern network requirements have diverged sharply. Real-time systems (e.g., High Frequency Trading) demand ultra-low latency with minimal logic delay, while hyperscale AI training requires maximum throughput for synchronized bulk transfers~\cite{HPCC}. A fundamental hardware trade-off makes it difficult for a single switch architecture to serve both. Increasing switch capability, such as adding deep buffering or complex scheduling, inevitably increases logic latency and resource usage, while general-purpose switches using fixed micro-architectures fail to adapt to diverse workloads. This architectural rigidity leaves significant performance on the table. As shown in Figure~\ref{fig:motivation} (left), switch architectures are highly sensitive to traffic patterns: with iSLIP-based design~\cite{islip} favoring uniform traffic and EDRRM-based design~\cite{li2002dual} handling bursts better. Therefore, it is essential to customize the switch architecture for different scenarios.

\looseness=-1
Beyond hardware logic, the transport protocol itself also matters. General-purpose protocols (e.g., Ethernet/IP) often impose unnecessary headers and processing overheads for specialized workloads. Commodity devices rigidly couple fixed hardware with standard parsers, lacking the flexibility to strip away distinct protocol layers or adjust logic for custom flows. While the emergence of P4~\cite{P4} allows for protocol customization in switches, their Protocol Independent Switch Architecture (PISA) designs limit the feasibility of improving performance by tweaking the switch architecture. As shown in Figure~\ref{fig:motivation} (right), by removing generic protocol overheads and tailoring the pipeline to the flow, the custom design achieves higher throughput. These results confirm that true optimization requires co-designing both the switch architecture and the protocol.

However, developing such co-design custom systems remains prohibitively difficult due to three critical challenges. 

\textbf{Challenge-1: Fixed Architectures versus Diverse Applications.} 
Co-optimizing the switch architecture with the protocol is essential, yet current solutions either lack architectural reconfigurability or remain tightly coupled to standard protocols. For instance, emerging applications often require specialized processing logic beyond standard forwarding, developers may need to inject custom compute kernels alongside custom protocols to maximize performance. However, fixed architectures (e.g., PISA) cannot handle complex stateful logic (e.g., floating-point aggregation). 

\textbf{Challenge-2: Efficient Simulation for Custom Protocol Network Switches.} An application's specific features can cause the switch to perform below its theoretical potential. Simulation in real-world traffic is crucial to identify bottlenecks, but current software simulations (e.g., {ns-3}~\cite{ns3}) lack native support for custom protocols and custom switch devices.

\textbf{Challenge-3: The Coupling Trap of Vertical Integration.} 
Optimizing a network for a specific application requires tight coupling between the protocol format and the switch micro-architecture. A small change in the application layer can ripple across the entire hardware stack, requiring developers to manually rewrite RTL parsing logic and re-tune architectural parameters. This cross-layer dependency makes it practically impossible to keep the network hardware synchronized with agile software iterations.

To address these challenges, we introduce \textbf{\sysname}, Switch and Protocol Adaptive Customization. \sysname bridges the gap between protocol definition and line-rate hardware generation using an HLS-based dynamic configurable switch template with hardware-calibrated design space exploration (DSE), enabling the automatic deployment of optimized networking substrates tailored for both performance and resources. 
Figure~\ref{fig:system} presents the end-to-end workflow. 

Our main contributions are as follows.
\begin{itemize}
    \item \textbf{Adaptive Network Switch Architecture:} To solve \textbf{Challenge-1}, we designed a high-performance modular HLS-based switch design that provides a custom-protocol parser, multiple forwarding structures, scheduling algorithms, and buffer allocation strategies. \sysname switch also provides the architectural hooks and interfaces to enable custom logic injection.
    
    \item \textbf{Efficient Custom Network-Stack Simulator:} To address \textbf{Challenge-2}, we created a multi-granularity network simulation system supporting both statistical and hardware-aware modeling. It enables the verification of generated switches and custom protocols on ns-3~\cite{ns3}, providing rapid performance insights prior to synthesis.

    \item \textbf{\sysname Domain-Specific Language (DSL) and Two-Stage Custom Network Workflow:} 
    We propose a two-stage workflow based on the \sysname DSL to streamline the development of programmable switches and tackle \textbf{Challenge-3}. By generating protocol drivers and HLS parsing libraries that link to switch behaviors via semantic binding, \sysname effectively decouples protocol specifications from the core switching logic. Furthermore, the framework enables automated DSE by integrating multi-granularity simulation with a hardware resource model. This allows for the rapid identification of Pareto-optimal designs, balancing latency and throughput trade-offs under strict hardware constraints.
\end{itemize}

\section{Background and Related Work}%

\subsection{Programmable Network Devices}
A network switch comprises four fundamental components: a packet parser, a forward table, packet buffers, and a switching fabric (scheduling algorithms). Existing FPGA-based research typically focuses on optimizing these individual modules in isolation. For instance, platforms like GCQ~\cite{10.1145/2145694.2145706}, 
Hipernetch~\cite{hipernetch}, SMiSLIP~\cite{meng2019investigating}, NetFPGA~\cite{netfpgasume} and CusComNet~\cite{CusCosNet} optimize the reconfigurable fabric layout. 
Other works focus on specific sub-problems, such as adapting iterative scheduling algorithms like iSLIP~\cite{islip} to meet FPGA timing constraints~\cite{papaphilippou2023experimental, meng2019investigating}. 
While techniques like hierarchical design~\cite{10.1145/2145694.2145706} and resource sharing improve local performance, these efforts are usually specific to particular architectures, lack a unified framework that allows developers to flexibly compose these optimized modules into a complete, customized switch.

\looseness=-1
A separate line of work focuses on programmable NIC and SmartNIC platforms. OpenNIC~\cite{open-nic} and Corundum~\cite{corundum} provide open-source FPGA NIC shells, while commercial DPUs (e.g., NVIDIA BlueField) offer fixed offload pipelines. eBPF-based approaches such as eBPFlow~\cite{ebpf_tnet} and Nanotube~\cite{nanotube} enable programmable packet processing on FPGA NICs within fixed pipeline structures. PANIC~\cite{panic} explores flexible multi-tenant NIC architectures, Pigasus~\cite{osdi20_zhao} demonstrates FPGA-based network security at 100\,Gbps, and FlowBlaze~\cite{flowblaze} introduces stateful packet processing abstractions. Recent efforts also investigate SmartNIC datapath acceleration~\cite{icnp2024}, software packet processing on FPGA NICs~\cite{hxdp}, and hardware offloading for virtual switches~\cite{sigcomm2024}. These systems primarily accelerate \textit{endpoint} functions (e.g., host offloads, protocol termination) rather than switch-level contention and queueing under multi-host traffic, which is the focus of \sysname.

\looseness=-1
In parallel to structural optimizations, researchers actively explore in-Network computing (INC) to offload application-specific computation. FPGAs serve as an ideal platform for this domain due to their flexibility~\cite{fahmy2025fpgas}. Recent studies demonstrate effective acceleration for distributed workloads, such as gradient aggregation for machine learning~\cite{li2019accelerating} and dynamic computation offloading~\cite{tokusashi2019case}. However, these implementations often operate as isolated point solutions. They typically embed custom logic within rigid pipelines or require significant manual effort to comply with infrastructure limitations~\cite{sapio2017network, kianpisheh2022survey}. This tight coupling compels developers to manually re-implement basic switching infrastructure for every new application~\cite{nickel2024survey}.
This limitation underscores the critical need for a modular, fully synthesizable switch architecture that can rapidly integrate custom computing kernels when application needs.

\subsection{Custom Network Protocols}
While Ethernet and TCP/IP~\cite{ethernet, rfc793} provide broad interoperability, their fixed header structures introduce substantial redundancy. Consequently, modern high-performance domains have adopted specialized protocols to meet stringent targets. For instance, RoCEv2~\cite{rocev2} for AI gradient synchronization, FIX~\cite{FIX50SP2} and FAST~\cite{FAST11} for ultra-low-latency trading, PROFINET~\cite{PROFINET} and EtherCAT~\cite{EtherCAT} for industrial control, and DCTCP~\cite{DBLP:conf/sigcomm/AlizadehGMPPPSS10} for data center traffic. Custom protocols can bring significant benefits to special scenarios. For example, underwater acoustic sensor networks often transmit payloads as small as 2 bytes~\cite{netblocks}. 
Wrapping these in standard headers results in a standard protocol that would take at least 42B, severely limiting goodput.

\subsection{DSLs for Network}
There are many DSLs to facilitate custom protocol development. For example, P4~\cite{P4} is a data-plane language supported by some commercial switches. Compilers such as VitisNetP4~\cite{vitisnet}, P4THLS~\cite{abbasmollaei2025p4thls}, P4-to-FPGA~\cite{cao2020p4}, and earlier efforts~\cite{p4_fccm2016, p4_sosr2017, p4_netfpga_wf} map P4 descriptions to FPGA logic, while configurable parser architectures~\cite{cabal_parser} target wire-speed throughput for arbitrary protocols. However, these tools primarily target data-plane parsing and match-action pipelines rather than complete switch architectures, and their reliance on fixed-stage pipeline structures limits global module configurability and stateful in-network computing capabilities. Another example is NetBlocks~\cite{netblocks}, which focuses on minimizing the overhead of custom protocol layout and generate high-performance system drivers. However, a software stack alone remains insufficient, as comprehensive protocol design requires validation through large-scale simulations and underlying hardware support.

\section{\sysname Overview and System}
\label{sec:overview}
\begin{figure*}[t]
    \centering
    \includegraphics[width=\linewidth]{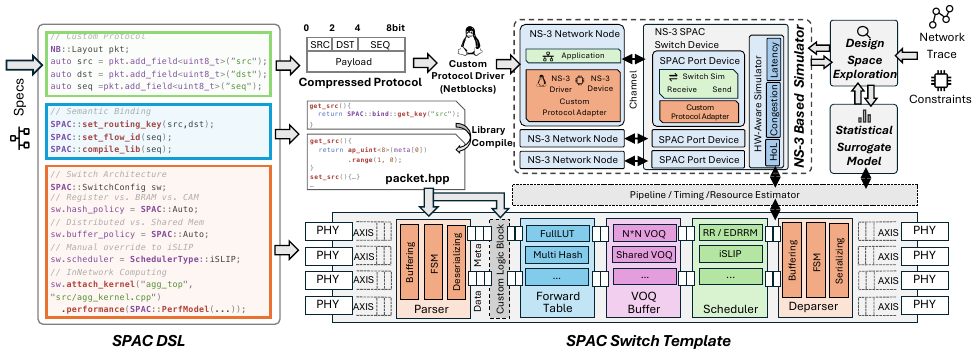}
    \vspace{-0.3in}
    \caption{\sysname System Overview.}
    \vspace{-0.1in}
    \label{fig:system}
\end{figure*}

Figure~\ref{fig:system} presents the overall architecture of \sysname, which comprises three key components.

\myparagraph{\sysname Switch Template.}
We abstract the switch architecture into 6 parts: \textit{Parser} (ingress and deserialize bitstreams), \textit{Custom Kernels}, \textit{Forward Table} (ingress and forwarding table lookup), \textit{VOQ Buffer} (buffer management and organization), \textit{Scheduler} (schedule different traffic to prevent HoL or packet loss) and \textit{Deparser} (egress and serialize packet). \sysname provides diverse hardware modules for each stage that adhere to a unified \textit{Meta+Data} I/O stream interface, enabling modular design composition tailored to specific application and throughput requirements (Section~\ref{sec:switch_detail}).

\myparagraph{NS-3 Based and Statistical Network Simulator.} Existing large-scale network simulators primarily support standard protocols and lack native support for custom protocols. 
We introduce multi-level simulation specifically designed to model custom protocol systems (Section \ref{sec:modeling}). The \textit{NS-3 Based Simulator} integrates our custom protocol adaptation layer into the NS-3 backend, and adds a \sysname switch device to provide more realistic network simulation. Meanwhile, the \textit{Statistical Surrogate Model} simulates packet arrival based on switch pipelines, end-to-end latency, and buffering to quickly obtain theoretical results.

\myparagraph{\sysname DSL and DSE Optimizer.} 
The \sysname DSL allows high-level descriptions of custom packet structures and device behavior, enabling users to focus on application-layer design without low-level hardware details. To rapidly explore a vast design space, our DSE employs multi-level simulation models to evaluate trade-offs between traffic patterns, device scale, and logic latency. This process identifies Pareto-optimal configurations and generates the final hardware and software stack (Section \ref{sec:dse}).

\subsection{The \sysname Specification Language}

Unlike prior approaches that require disjoint specifications for software drivers, hardware logic, and simulation models, \sysname serves as a single source of truth. The DSL decouples the logical definition of the network protocol from its physical realization in the FPGA fabric, enabling automated DSE. As shown in Figure~\ref{fig:system} (left), the language primitives are categorized into three abstraction layers: Custom Protocol Definition, Semantic Binding, and Architecture Configuration.

\myparagraph{Custom Protocol Definition.} 
\sysname uses NetBlocks-compatible syntax\cite{netblocks} to specify custom protocols. By supporting bit-level serialization, NetBlocks facilitates the generation of highly optimized, compressed protocols. Additionally, by integrating the underlying high-performance network driver stack, it offers a stable ABI across simulation and real-world deployment, effectively streamlining the development workflow.

\myparagraph{Semantic Binding.}
Each user-defined protocol bit-field has a semantic alias, which is mapped to the switch architecture during  Semantic Binding. As shown in Figure~\ref{fig:system}, the protocol field designated for routing (i.e., the \texttt{routing\_key}) must be specified, while other fields remain optional. During the header compilation stage, the \sysname Compiler locates these fields within the protocol via key-value matching and generates inlined parsing logic into the HLS header file (\texttt{packet.hpp}) for both the user kernel and the switch code.
This decoupling allows users to modify the protocol layout without altering the underlying HLS implementation. By leveraging template metaprogramming, the system efficiently generates hardwired logic during switch synthesis, enabling line-rate packet parsing.

\myparagraph{Architecture Configuration.}
A key innovation of \sysname is its support for different switch architectures. Instead of forcing developers to write rigid HLS code, the DSL allows the specification of architectural policies (e.g., BufferPolicy, HashPolicy) as either explicit values or Auto. When policy is set to Auto, the \sysname DSE engine infers the optimal micro-architecture selections based on traffic characteristics and network topology.
To support application-specific logic beyond standard forwarding, \sysname also provides an injection point for custom HLS kernels. To incorporate these kernels into the DSE loop, users specify latency and resource boundaries via a \texttt{performance} interface.

\lstset{
  language=C++,
  basicstyle=\footnotesize\ttfamily,
  breaklines=true,
  breakatwhitespace=false,
  captionpos=b,
  extendedchars=false,
  frame=single,
  framerule=0pt,
  keywordstyle=\color{blue}\bfseries,
  commentstyle=\color{olive},
  stringstyle=\color{purple},
  numbers=left,
  numbersep=5pt,
  showstringspaces=false,
  showspaces=false,
  showtabs=false,
  tabsize=4
}

\subsection{Configurable Switch Architecture}
\label{sec:switch_detail}

To support the combination of diverse network fabrics, the \sysname switch design adopts a decoupled, stream-based modular design. We standardize inter-module communication using a dual-stream interface: a standard AXI-Stream for payload propagation and a compiler-synthesized MetaData side-channel for header (e.g., extracted routing keys, QoS). 
By encapsulating architectural state within the MetaData stream, the design ensures strict isolation between stages. For instance, the \textit{Forward Table} can be seamlessly swapped between a FullLookUpTable array and a MultiBank Hash table without affecting the downstream Scheduler, while custom computing kernels can be injected post-parsing with zero glue logic. 

\subsubsection{Protocol-Aware Parser}
\looseness=-1
To achieve line-rate processing, \sysname abandons the traditional ``programmable parser" architecture (which often relies on TCAMs or runtime configuration registers) in favor of a template-driven synthesis approach. The core of our parsing subsystem is a generic HLS template, which remains completely agnostic to specific protocol details until compilation. During the synthesis phase, the \sysname compiler lowers the high-level user-defined protocol specification into a C++ header. By instantiating the HLS parser template with these generated traits, we use C++ metaprogrammming to recursively compute the exact bit-offset of every field relative to the AXI-Stream flit boundaries at compile time. This mechanism automatically detects fields that straddle word boundaries to synthesize minimal state retention logic only when strictly necessary, while lowering intra-flit field accesses into hard-wired bit-slicing operations. Consequently, the resulting hardware achieves efficiency comparable to hand-optimized RTL while retaining the flexibility of high-level definitions.

\subsubsection{Forward Table}
\begin{figure}[tbp]
  \centering 
  \includegraphics[width=0.9\linewidth]{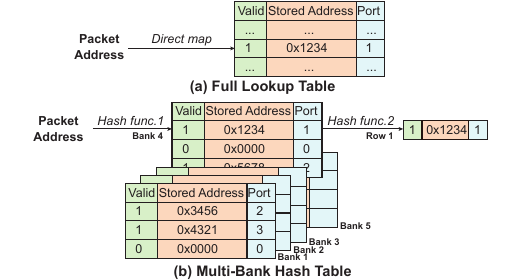}
  \caption{Forward Table Architectures}
  \label{fig:multibank}
  \vspace{-0.1in}
\end{figure}

The Forward Table maintains address-port mappings: it queries the destination address to determine the output port (or broadcast) and learns the source address on every arrival. The core optimization lies in choosing a data structure that enables multiple ports to access the table in minimal clock cycles. We implement two forwarding table variants:

The array-based \textbf{Full Lookup Table} utilizes a one-dimensional table where the address field serves as the direct index. We fully partitioned the table, and it can support simultaneous reads and writes from multiple ports in a single cycle. While highly efficient and logic-light for short addresses, such as NetBlocks’ shrunk protocol, which is used in underwater robots communication, it is unsuitable for long addresses as memory usage increases exponentially.
To handle longer address fields, we also provide \textbf{Multi-Bank Hash Table}, as shown in Figure~\ref{fig:multibank}, which uses two-dimensional tables with hash functions for indexing. The table is partitioned into multiple banks so that each port’s input ideally maps to a distinct bank, minimizing conflicts. Although this allows for larger address spaces, it introduces tradeoffs in the form of additional logic for hash calculations and conflict resolution.

\subsubsection{Virtual-Output-Queue Buffer}
\begin{figure}[tbp]
  \centering
  \includegraphics[width=0.9\linewidth]{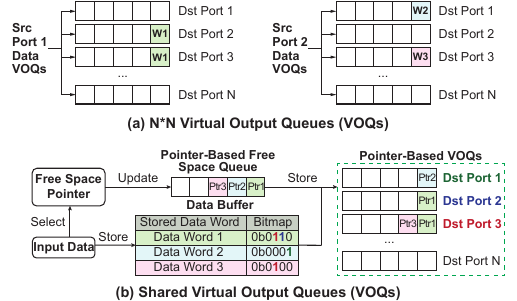}
  \caption{N*N Virtual Output Queues (VOQs) versus Shared VOQs.}
  \label{fig:VOQ}
  \vspace{-0.2in}
\end{figure}

The \textit{VOQ Buffer} is vital for decoupling input and output traffic. After the packet’s destination port is labeled, it is stored into FIFO data queues based on source and destination port information. To achieve high throughput, each port maintains its own data queue to process packets in parallel. However, a single queue suffers from Head-of-Line (HoL) blocking, where a congested destination port prevents other ports from being served. To address this, we design two type of Virtual Output Queues (VOQs), as illustrated in Figure~\ref{fig:VOQ}. 

\looseness=-1
\textbf{N*N Data VOQs} is straightforward: packets labeled with a destination port go to the corresponding queue; broadcast packets are copied and stored in all queues associated with the source port. The queues are fully partitioned to support reading and writing in parallel. One drawback of this implementation is that it suffers from limited FIFO depth and high memory/time costs due to data duplication for broadcasting. As a compromise, we also provide \textbf{Shared VOQs}~\cite{meng2019investigating}, which use a central data buffer with pointer-based queues: instead of copying packet data, it stores each packet once and replicates only its pointer, using a bitmap to track pending destinations. This offers memory efficiency but introduces logic overhead for pointer management, which may impact performance.

\subsubsection{Scheduler}
The scheduling algorithm is critical for arbitrating access between input and output ports. Its primary objectives are to maximize throughput by optimizing the number of input-output matchings per cycle and to ensure fairness, preventing any source or destination port from starvation. \sysname supports three scheduling architectures:
\textbf{Round-Robin} uses a simple cyclic priority rotation. It is hardware-efficient but its combinational logic latency grows with port count, requiring up to $N$ cycles in the worst case. \textbf{iSLIP}~\cite{islip} employs an iterative three-phase (\textit{Request}, \textit{Grant}, \textit{Accept}) process with independent rotating pointers. While it theoretically achieves 100\% throughput, the ``Find-First'' operation creates long combinatorial paths that often become the FPGA timing bottleneck. To address this, \textbf{EDRRM} simplifies the matching to a two-phase \textit{Request} and \textit{Grant} process combined with an exhaustive service strategy, reducing arbitration overhead while maintaining high efficiency for bursty traffic.

\begin{figure}[t]
    \centering
    \includegraphics[width=248pt]{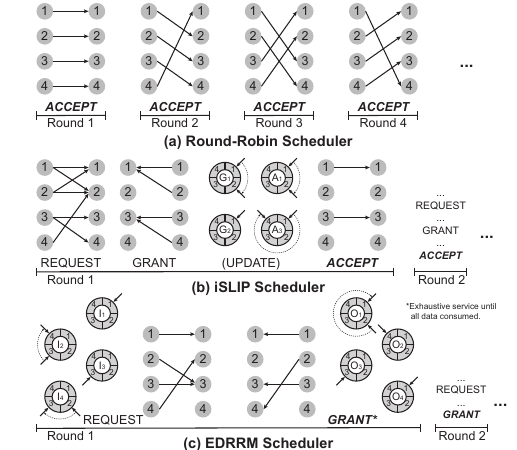}
    \caption{RR / EDRRM / iSLIP Scheduler.}
    \label{fig:islip}
    \vspace{-0.2in}
\end{figure}

\subsubsection{Support for User-Defined Logic}
\sysname supports extensibility through a modular architecture paired with an exported HLS protocol header library. This empowers users to link custom kernels directly into the switch pipeline, utilizing our library to simplify protocol parsing. This combination lays a concrete foundation for future in-network computing research by significantly lowering the development barrier for custom hardware acceleration.

\section{Design Space Exploration}
\sysname comes with a system of tools for enumerating the custom protocol design space. They allow for the efficient simulation of custom networks, the discovery of the optimal switch architecture for a given protocol and finally RTL accurate network simulations. This system is designed to enable rapid experimentation with custom protocols and switch architectures, before commitment to compute-intensive cycle-accurate analysis is required.

\subsection{Multi-level Simulation Model}
\label{sec:modeling}

\subsubsection{Hardware-Aware ns-3 Based Switch Simulator}

We present a flexible switch simulation architecture built upon the well-known NS-3 backend. The NS-3 network simulator contains four-layer network node abstractions: application layer, host network stack layer, Ethernet device layer, and channel layer. As shown in Figure~\ref{fig:system}, by abstracting the conventional Ethernet layer into a generalized \textit{Custom Protocol Adapter}, our design seamlessly integrates DSL-compiled drivers (handling logic such as parsing and retransmission) thereby transcending the limitations of standard Ethernet models. This architecture encapsulates protocol state to enable multi-instance concurrency, ensuring fidelity to the specification.

On the switch side, \sysname employs a Hardware-Aligned Modeling approach. We abstract switch ports as specialized NS-3 nodes (SPAC Port Device) and explicitly model internal behaviors such as forwarding table lookups, packet buffering, and scheduling in a unified simulator. A distinguishing feature of \sysname is its support for \textbf{Hardware Back-Annotation}: by injecting performance metrics from physical FPGA experiments into the model as processing parameters, we ensure the simulator reflects realistic hardware constraints. Users can choose to enable back-annotation for high-fidelity evaluation of latency and resources, or disable it for rapid functional testing, thus balancing simulation accuracy with performance.

\subsubsection{Statistical Simulator}
To circumvent the prohibitive cost of cycle-accurate simulations for large-scale traces, we incorporate a \textbf{Statistical Surrogate Model} within the DSE loop. This model exploits the inherent determinism of FPGA-based switching logic. Specifically, the fixed Initiation Interval (II) and predictable pipeline latency of HLS modules. Instead of simulating signal-level transitions, we abstract the switch datapath into a lightweight, event-driven transaction model. 
By parameterizing this model with static hardware attributes (e.g., bus width, arbitration latency, and pipeline depth), our engine can process traces in seconds.

This approach enables a rapid yet accurate estimation of critical micro-architectural metrics prior to synthesis. The model estimates line-rate feasibility, BRAM lower bounds from peak VOQ occupancy, and latency distributions by combining deterministic pipeline delays with dynamic queuing effects.

\begin{figure}[t]
    \centering
    \includegraphics[width=\linewidth]{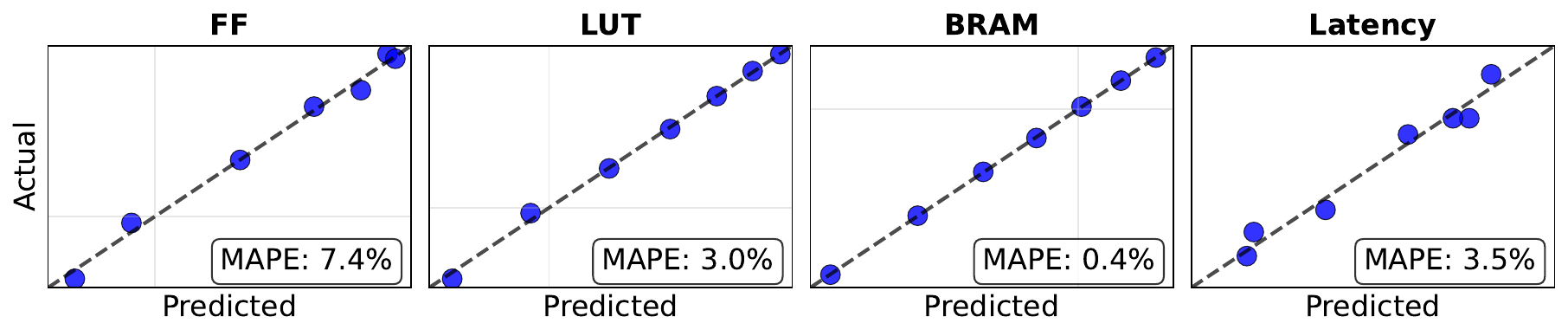}
    \vspace{-0.3in}
    \caption{Variance Analysis of Resource and Performance Estimates}
    \label{fig:estimate-resource}
    \vspace{-0.15in}
\end{figure}
To validate the fidelity of this surrogate model, we cross-verified its predictions against post-synthesis timing reports from Vitis HLS. Figure~\ref{fig:estimate-resource} shows our surrogate model and real-world design evaluation of the resource and latency aspects of a 2-8 port design. The result with Mean Absolute Percentage Error (MAPE) of 0.4\%$\thicksim$7.4\% confirms that our lightweight abstraction effectively captures the micro-architectural impact, providing a trustworthy basis for the DSE engine.

\vspace{-0.05in}
\subsection{DSE Algorithm}
\label{sec:dse}
To navigate the design space efficiently, \sysname adopts a \textit{Progressive Constraint Satisfaction} strategy. It can gradually increase the simulation granularity while reducing the search space. Algorithm \ref{alg:dse_progressive} details this process.

The DSE engine characterizes the input trace $\mathcal{T}$ into a feature vector $\mathbf{f} = [\mathcal{I}_{burst}, \mathcal{H}_{addr}, S_{min}]$. Where $\mathcal{I}_{burst}$ shows the Index of Dispersion for Counts (IDC), acting as a congestion proxy. $\mathcal{H}_{addr}$ is the entropy of destination addresses, indicating the effectiveness of caching, and $S_{min}$ shows the minimum packet payload observed in the windowed trace. This metric defines the worst-case arrival rate and sets the strict timing budget for the pipeline.

\looseness=-1
We apply hardware constraints with tolerance margins to prune architecturally infeasible candidates while preserving designs that may recover performance through compiler optimizations or resource over-provisioning. For an architecture template $a$ with an initiation interval $II_a$, we introduce a timing relaxation factor $\delta$ and make sure $T_{proc} > (1 + \delta) \cdot T_{arrival}$, where $T_{proc} = II_a/F_{clk}$ and $T_{arrival} = (S_{min} \times 8)/\text{LinkRate}$.
For the surviving candidates, we execute a One-Shot \textit{Ideal} Simulation with infinite buffer constraints. We record the maximum queue occupancy ($Q_{max}$) and the latency distribution for every port in the trace. If a design point violates the 99th-percentile latency SLA even with infinite buffering, it will be dropped.
Next, we explore the optimal VOQ buffer size based on statistical guarantees. Using the queue occupancy histogram collected in Stage 2, we identify the specific queue depth $d_{opt}$ corresponding to the target tail drop rate $\epsilon$. We then map $d_{opt}$ to physical FPGA resources by aligning it with the switch data width. Candidates that violate the total BRAM capacity constraint are pruned. 
Finally, the engine runs a verification simulation with the derived parameters to ensure the discretized sizing meets all SLAs.
\begin{figure}[t]
    \centering
    \includegraphics[width=0.8\linewidth]{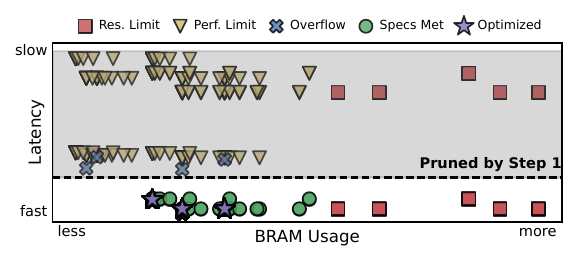}
    \vspace{-0.2in}
    \caption{DSE Algorithm Search Space Visualization.}
    \vspace{-0.1in}
    \label{fig:dse_space}
\end{figure}

{
\removelastskip
\SetAlgoSkip{empty}
\setlength{\intextsep}{0pt}
\setlength{\textfloatsep}{0pt}
\begin{algorithm}[t]
\SetNlSty{tiny}{}{}         %
    \SetNlSkip{0.5em}           %
    \small                      %
    \linespread{0.85}\selectfont %
    \small
    \caption{Progressive Constraint Satisfaction DSE}
    \label{alg:dse_progressive}
    
    \SetKwInOut{Input}{Input}
    \SetKwInOut{Output}{Output}
    \Input{Trace $\mathcal{T}$, Templates $\mathcal{A}$, Constraints $\mathcal{C}_{SLA}, \mathcal{C}_{Res}$}
    \Output{Optimal Configuration $\mathbf{x}^*$}
    \BlankLine 

    \tcp{Stage 1: Static Pruning}
    $\mathbf{f} \leftarrow [\mathcal{I}_{burst}, \mathcal{H}_{addr}, S_{min}] \leftarrow \text{Analyze}(\mathcal{T})$ \;
    
    $\mathcal{A}_{active} \leftarrow \mathcal{A}$ \;
    
    \ForEach{$a \in \mathcal{A}_{active}$}{
        $T_{arrival} \leftarrow (S_{min} * 8) / \text{LinkRate}$ \;
        
        $T_{proc} \leftarrow a.\text{II} / \text{Freq}$ \;
        
        \If{$T_{proc} > (1 + \delta) * T_{arrival}$}{
            $\mathcal{A}_{active}.\text{remove}(a)$ \;
        }
    }

    \tcp{Stage 2: Coarse-grained Profiling}
    $\mathcal{A}_{valid} \leftarrow \emptyset$ \;
    
    \ForEach{$a \in \mathcal{A}_{active}$}{
        \tcp{Infinite Buffer Simulation}
        $Q_{hist}, L_{dist} \leftarrow \text{Run\_Surrogate\_Model}(a, \mathcal{T})$ \;
        
        \If{$\text{Percentile}(L_{dist}, 99) \le \mathcal{C}_{SLA}.\text{Latency}$}{
            $\mathcal{A}_{valid}.\text{add}(\{a, Q_{hist}, L_{dist}\})$ \;
        }
    }

    \tcp{Stage 3: Statistical Sizing}
    $\mathbf{x}^* \leftarrow \text{NULL}$ \;
    
    \ForEach{$\{a, Q_{hist}\} \in TopKLatency(\mathcal{A}_{valid})$}{
        $d_{aligned} \leftarrow \text{AlignToBRAM}(Q_{hist}, \text{Width})$ \;
        
        \If{$\text{TotalBRAM}(d_{align}) \le \mathcal{C}_{Res}$}{
            $\mathbf{x}_{curr} \leftarrow \{a, d_{aligned}\}$ \;
            
            \If{$\text{Run\_NS3\_Sim}(\mathbf{x}_{curr}) \text{ meets } \mathcal{C}_{SLA}$}{
                $\text{UpdateOptimal}(\mathbf{x}^*, \mathbf{x}_{curr})$ \;
            }
        }
    }

    \Return{$\mathbf{x}^*$}
\end{algorithm}
}

\looseness=-1
To validate, we designed a brute-force enumeration. Figure~\ref{fig:dse_space} plots the primary resource usage (BRAM) and latency for Incast small-packet bursts across all combinations of architectures and buffer sizes. The design points identified by \sysname DSE (marked with $\bigstar$) lie on the Pareto-optimal frontier: the first stage prunes infeasible architectures, and the trace-aware buffer allocation then locates the resource-minimal solution.

\section{Evaluation}
\begin{table*}[t]
\centering
\resizebox{0.95\linewidth}{!}{
\begin{threeparttable}
\caption{Comparison of Related Work: Functions, Implementation, Resource Overhead, and Performance.}
\label{tab:eval-main}
\setlength{\tabcolsep}{4.5pt}

\begin{tabular}{lccccccccccccc}
\hline
\multicolumn{4}{c}{\textbf{Configuration}}                                                                                                                                                   & \multicolumn{4}{c}{\textbf{Function}}                                                                                                                             & \multicolumn{3}{c}{\textbf{Resources}}     & \multicolumn{3}{c}{\textbf{Performance}}                                                                                                                                                          \\ \hline
\textbf{Switch}                      & \textbf{\begin{tabular}[c]{@{}c@{}}Design\\ Language\end{tabular}} & \textbf{\begin{tabular}[c]{@{}c@{}}Width\\ (bits)\end{tabular}} & \textbf{Ports} & \textbf{Parser} & \textbf{\begin{tabular}[c]{@{}c@{}}Fwd\\ Table\end{tabular}} & \textbf{Scheduler} & \textbf{\begin{tabular}[c]{@{}c@{}}VOQ\\ Buffer\end{tabular}} & \textbf{\begin{tabular}[c]{@{}c@{}}LUT\\ (K)\end{tabular}} & \textbf{\begin{tabular}[c]{@{}c@{}}FF\\ (K)\end{tabular}} & \textbf{BRAM} & \textbf{\begin{tabular}[c]{@{}c@{}}Freq\\ (MHz)\end{tabular}} & \textbf{\begin{tabular}[c]{@{}c@{}}Latency\\ (ns)\end{tabular}} & \textbf{\begin{tabular}[c]{@{}c@{}}Max Throughput\\ (Gbps)\end{tabular}} \\ \hline
GCQ~\cite{10.1145/2145694.2145706}   & RTL                                                                & \begin{tabular}[c]{@{}c@{}}1024\\ (256)\end{tabular}            & 16             & \ding{56}              & \ding{56}                                                           & RR-Only\tnote{1}            & \ding{52}                                                            & 60.0         & 13.5        & 224           & 180                                                           & 172                                                             & \begin{tabular}[c]{@{}c@{}}160\\ (40.9)\end{tabular}                 \\ \hline
SMiSLIP~\cite{meng2019investigating}\tnote{2} & RTL                                                                & 256                                                             & 16             & \ding{56}              & \ding{56}                                                           & \ding{52}                 & \ding{52}                                                            & 80.3         & 23.1        & NA            & 140                                                           & 41.4                                                            & 35.8                                                                 \\
                                     &                                                                    & 512                                                             & 16             & \ding{56}              & \ding{56}                                                           & \ding{52}                 & \ding{52}                                                            & 123.8        & 24.5        & NA            & 110                                                           & 44.3                                                            & 56.3                                                                 \\ \hline
Hipernetch~\cite{hipernetch}\tnote{2}         & RTL                                                                & 256                                                             & 16             & \ding{56}              & \ding{56}                                                           & \ding{52}                 & \ding{56}                                                            & 150          & 125         & NA            & 225                                                           & 66                                                              & 58                                                                   \\
                                     &                                                                    & 512                                                             & 16             & \ding{56}              & \ding{56}                                                           & \ding{52}                 & \ding{56}                                                            & 200          & 300         & NA            & 175                                                           & 87                                                              & 89                                                                   \\ \hline
VitisNetP4 \cite{vitisnet}                & P4+RTL                                                             & 256                                                             & -              & \ding{52}              & \ding{52}                                                           & \ding{56}                 & \ding{56}                                                            & 5.21         & 8.99        & 9             & 259                                                           & 176                                                             & 66.3                                                                 \\ \hline

P4-to-FPGA \cite{cao2020p4}          & P4+RTL                                                             & 272                                                             & -              & \ding{52}              & \ding{52}                                                           & \ding{56}                 & \ding{56}                                                            & 5.64         & 1.94        & 2             & 176                                                           & 40                                                              & 48.1                                                                 \\ \hline
P4THLS \cite{abbasmollaei2025p4thls} & P4+HLS                                                             & 256                                                             & -              & \ding{52}              & \ding{52}                                                           & RR-Only\tnote{1}            & \ding{56}                                                            & 5.97         & 3.58        & 4             & 250                                                           & 40                                                              & 69.5                                                                 \\ \hline
\textbf{SPAC Core-Only}              & HLS                                                                & 256                                                             & -              & \ding{52}              & \ding{52}                                                           & RR-Only\tnote{1}            & \ding{56}                                                            & 4.47         & 7.01        & 2             & 350                                                           & 34.3                                                            & 89.6                                                                 \\ \hline
\textbf{SPAC Ethernet}               & HLS                                                                & 512                                                             & 8              & \ding{52}              & \ding{52}                                                           & \ding{52}                 & \ding{52}                                                            & 80.1         & 45.8        & 304           & 146                                                           & 68.3                                                            & 74.7                                                                 \\
                                     &                                                                    &                                                                 & 16             & \ding{52}              & \ding{52}                                                           & \ding{52}                 & \ding{52}                                                            & 315.6        & 135.1       & 608           & 137                                                           & 109.2                                                           & 70.1                                                                 \\ \hline
\textbf{SPAC Basic}                  & HLS                                                                & 256                                                             & 8              & \ding{52}              & \ding{52}                                                           & \ding{52}                 & \ding{52}                                                            & 38.9         & 30.5        & 260           & 165                                                           & 57.3                                                            & 84.5                                                                 \\
                                     &                                                                    &                                                                 & 16             & \ding{52}              & \ding{52}                                                           & \ding{52}                 & \ding{52}                                                            & 96.1         & 83.2        & 498           & 142                                                           & 85.5                                                            & 72.7                                                                 \\ \hline
\end{tabular}

\begin{tablenotes}
      \footnotesize
      \item[1] Only support Round-Robin scheduler.
      \item[2] BRAM usage not provided.
    \end{tablenotes}
\end{threeparttable}
}
\vspace{-0.2in}
\end{table*}

\subsection{Experiment Setup}
\myparagraph{Hardware.}
We implement our design using Vitis HLS 2023.2 and synthesize for an AMD Alveo U45N Network Accelerator Card (xcu26-vsva1365-2LV-e), hosted on a server equipped with an AMD EPYC 9335 32-core processor. The target clock period is set to 350MHz. 

\myparagraph{Workloads.}
We employ multiple traffic traces:
\begin{itemize}
    \item Industry: Sources real-world SCADA operations from the Dataset of SCADA traffic captures from a medical waste incinerator~\cite{SCADA}.
    \item High-Frequency Trading (HFT): generated based on traffic patterns from a real-world HFT Company~\cite{NFIM}, featuring low latency and high burstiness.
    \item RL(All-Reduce): Simulates traces in distributed reinforcement learning based on iSwitch~\cite{iswitch}, showing regularity and hotspots.
    \item DataCenter: Models microservice dependencies from the Alibaba Cluster Trace~\cite{alibaba} and randomly deploy them to 8 nodes to simulate Kubernetes-based containerization.
    \item Underwater: Simulates underwater communication among 8 underwater robots using the DESERT~\cite{DESERT} scheme, characterized by regular communication with minimal payloads ($\approx$2B).
\end{itemize}

\myparagraph{Baselines.}
Since most prior FPGA-based switch designs focus on individual modules (e.g., scheduler or crossbar only), direct end-to-end comparison with SPAC, which covers the complete switch pipeline (parser, forwarding table, scheduler, and VOQ buffer), is inherently limited. We therefore adopt \textbf{SPAC Ethernet}---Ethernet protocol with MultiBankHash, $N \times N$ VOQ, and iSLIP scheduling---as the default baseline, representing a general-purpose design point. We structure experiments into three levels: (1)~unloaded datapath comparison (Table~\ref{tab:eval-main}, post-place-and-route); (2)~scalability analysis under varying port counts (Figure~\ref{fig:perf_port_arch}); and (3)~trace-driven DSE co-optimization (Table~\ref{tab:app-compare}, hardware-aware ns-3 simulation with cycle-level back-annotation).

\subsection{Resource Efficiency \& Scalability}
\subsubsection{Resource Usage}
Table~\ref{tab:eval-main} reports unloaded datapath properties. Max Throughput is defined as $\text{datawidth} \times \text{II} \times f_{\text{max}}$; Latency refers to single-packet port-to-port traversal without contention.
We synthesized three configurations for comparison:
\textbf{SPAC Core-Only} retains only basic scheduling and parsing to isolate core datapath latency. \textbf{SPAC Ethernet} is the baseline (Section~6.1). \textbf{SPAC Basic} uses a compressed protocol for small-scale networks while keeping the same underlying architecture.

\looseness=-1
Most previous approaches use manually optimized RTL, offering deterministic latency at the cost of maintainability.
GCQ achieves 160 Gbps theoretical throughput via grouping, but its single-bus bandwidth is only 40.9 Gbps (vs. our 70.1 Gbps), and clock domain crossing introduces 172ns latency---1.57$\times$ our unoptimized and 2.01$\times$ our optimized design.
SMiSLIP uses shared buffering for lower resource utilization, but its centralized scheduler limits throughput as port count grows.
Hipernetch achieves high throughput via pipelined parallel round-robin arbiters, at the cost of significant resource consumption.

Note that these prior designs implement \textit{only} scheduler/crossbar modules, whereas SPAC includes the full-stack datapath (parsing, forwarding-table learning/lookup, scheduling, and buffering), which increases pipeline depth and end-to-end latency.
Although the resource consumption of SPAC Ethernet and SPAC Basic is slightly higher than these works, we achieve performance comparable to manually tuned RTL using HLS. More importantly, we provide a flexible, configurable, full-stack switch architecture, including automatic look-up table learning and protocol adaptation, which is ignored by other works.
To demonstrate the efficiency of our core design, we synthesized SPAC Core, retaining only the simplest scheduler and packet parsing logic, to compare against VitisNetP4, P4-to-FPGA and P4THLS, which also only include simple parsing and forwarding logic. Since the majority of SPAC's resources are typically dedicated to organizing complex schedulers and VOQ buffers, our streamlined design achieves lower LUT consumption and a 1.4$\thicksim$2.0$\times$ frequency improvement, thereby yielding higher theoretical throughput.

\subsubsection{Scalability}
\begin{figure}[t]
    \centering
    \includegraphics[width=\linewidth]{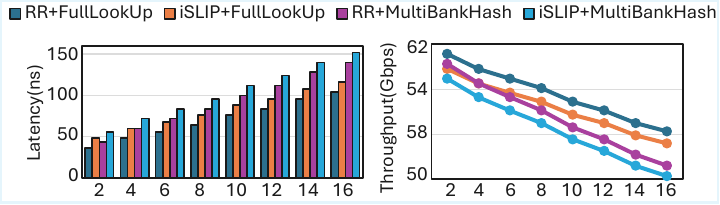}
    \vspace{-0.25in}
    \caption{Average P2P Performance Under Different \#Port and Architectures.}
    \vspace{-0.1in}
    \label{fig:perf_port_arch}
\end{figure}
\sysname supports diverse combinations of switch architectures and port configurations. Due to the inherent non-determinism of HLS synthesis, latency and throughput fluctuate as the port count changes. 
We show SPAC Ethernet's performance with medium-sized packets ($\approx$ 512B). As shown in Figure \ref{fig:perf_port_arch}, latency and throughput exhibit approximately linear trends as port counts rise. This behavior stems from the increasing complexity of scheduling decisions and the significant lookup delays introduced by the expansion of forwarding table capacity.
Under high loads, complex combinatorial logic increases the pipeline initiation interval, degrading average throughput. The decline rate correlates with forwarding table architecture, as MultiBankHash bank conflicts increase with lookup traffic.
Overall, our design achieves a latency of approximately 109ns at 16 ports, only 63.4\% of the latency observed in the GCQ switch with identical specifications, demonstrating the performance of our HLS-implemented modules.

\subsection{Domain Specific Adaption}
\begin{table}[]
\centering
\setlength{\tabcolsep}{0.5pt}
\renewcommand{\arraystretch}{1}
\caption{Comparison of switch design and performance after optimization in different applications}
\label{tab:app-compare}
\resizebox{0.95\linewidth}{!}{
\begin{threeparttable}
\begin{tabular}{lcccccc}
\hline
\textbf{\begin{tabular}[c]{@{}l@{}}Application\\ (Num nodes)\end{tabular}} & \textbf{Architecture}                                                   & \textbf{\begin{tabular}[c]{@{}c@{}}Data\\ Width\\ (Bits)\end{tabular}} & \textbf{\begin{tabular}[c]{@{}c@{}}Avg Header\\ (Payload)\\ (Bytes)\end{tabular}} & \textbf{\begin{tabular}[c]{@{}c@{}}Avg Opt\\ Latency\\ (ns)\end{tabular}} & \textbf{\begin{tabular}[c]{@{}c@{}}Baseline\\ Latency\\ (ns)\end{tabular}} & \textbf{\begin{tabular}[c]{@{}c@{}}Latency\\ Breakdown\end{tabular}} \\ \hline
\begin{tabular}[c]{@{}l@{}}HFT \\ (8)\end{tabular}                         & \begin{tabular}[c]{@{}c@{}}FullLookUp\\ N*N VOQs\\ RR\end{tabular}      & 256                                                                    & \begin{tabular}[c]{@{}c@{}}2\\ (24)\end{tabular}                                  & 64                                                                        & 103.9                                                                      & 38.4\%                                                               \\ \hline
\begin{tabular}[c]{@{}l@{}}RL\\ (8)\end{tabular}                           & \begin{tabular}[c]{@{}c@{}}FullLookUp\\ N*N VOQs\\ EDRRM\end{tabular}   & 1024                                                                   & \begin{tabular}[c]{@{}c@{}}2\\ (1463)\end{tabular}                                & 538                                                                       & 20391.7                                                                    & 97.3\%\tnote{$\dagger$}                                                                    \\ \hline
\begin{tabular}[c]{@{}l@{}}Datacenter\\ (32)\end{tabular}                  & \begin{tabular}[c]{@{}c@{}}MultiBank\\ Shared VOQs\\ iSLIP\end{tabular} & 256                                                                    & \begin{tabular}[c]{@{}c@{}}4\\ (965.5)\end{tabular}                               & 154.1                                                                     & 167.2                                                                      & 7.8\%                                                                \\ \hline
\begin{tabular}[c]{@{}l@{}}Industry\\ (10)\end{tabular}                    & \begin{tabular}[c]{@{}c@{}}FullLookUp\\ Shared VOQs\\ RR\end{tabular}   & 128                                                                    & \begin{tabular}[c]{@{}c@{}}2\\ (58.7)\end{tabular}                                & 76                                                                        & 119.9                                                                      & 36.6\%                                                               \\ \hline
\begin{tabular}[c]{@{}l@{}}Underwater\\ (8)\end{tabular}                   & \begin{tabular}[c]{@{}c@{}}FullLookUp\\ Shared VOQs\\ RR\end{tabular}   & 256                                                                    & \begin{tabular}[c]{@{}c@{}}2\\ (2)\end{tabular}                                   & 42                                                                        & 68.3                                                                       & 38.5\%                                                               \\ \hline
\end{tabular}
\begin{tablenotes}
\footnotesize
\item[$\dagger$] Baseline experienced packet loss during incast.
\end{tablenotes}
\end{threeparttable}
}
\vspace{-0.2in}
\end{table}

We conducted experiments on customized switch architectures using various network flows with distinct characteristics. These flows represent scenarios characterized by high-frequency small packets, ultra-low latency requirements, high-throughput incast patterns, and large-scale sparse connectivity. Table~\ref{tab:app-compare} presents the switch architectures and packet statistics resulting from the DSE trade-offs.

For small-scale networks, \sysname reduces protocol overhead via header compression (14B to 2B), enabling the use of low-latency FullLookup forwarding tables. In small-packet environments like HFT, the algorithm selects pipelined Round-Robin scheduling to achieve higher frequency and maintain line-rate throughput, whereas iSLIP is preferred for DataCenter workloads to mitigate HoL blocking under mixed traffic, providing lower average latency. The RL workload tends to converge on EDRRM, as it effectively balances the handling of bursty traffic with latency constraints. For sensor networks such as underwater robots, we  compress the packet to just 4B~\cite{netblocks}. This allows for a minimalist architecture with simplified buffering and scheduling logic. Compared to the SPAC Ethernet baseline at the same 8-port scale, this reduces LUT and BRAM usage by $\thicksim$$55\%$ and $\thicksim$$53\%$ respectively, while lowering latency to only 42ns.

While both DataCenter and Industrial scenarios opted for Shared VOQs, the underlying motivations differ. In DataCenter environments, the network scale typically involves a large number of connected nodes ($N$). Implementing fully partitioned $N \times N$ VOQs results in quadratic resource complexity ($O(N^2)$), which imposes excessive pressure on limited FPGA BRAM resources. Unlike HFT scenarios that demand ultra-low, deterministic latency, DataCenter workloads generally exhibit looser latency constraints, allowing them to tolerate the slight logic overhead of pointer management. Given that RPC traffic in data centers consists primarily of unevenly distributed mice flows~\cite{DBLP:conf/sigcomm/RoyZBPS15}, Shared VOQs provide superior buffer utilization efficiency to absorb bursts without the prohibitive resource costs of static partitioning.

These results reveal how traffic characteristics drive architectural choices. In small-packet latency-sensitive workloads such as HFT, frequency dominates over scheduling fairness, so DSE favors pipelined RR to maximize $f_{\text{max}}$. Under bursty incast patterns like RL All-Reduce, buffering and scheduling become the dominant factors for tail latency beyond line-rate, leading DSE to select wider buses with EDRRM. At larger scale in DataCenter workloads, $O(N^2)$ VOQ resource pressure makes buffer organization the primary bottleneck, and DSE shifts to shared VOQs with iSLIP to balance HoL blocking mitigation against resource constraints.

Our DSE model also optimized the switch bus width. In most small-scale, high-frequency scenarios, a 256-bit bus width provides sufficient bandwidth, making further expansion a resource waste. However, for bandwidth-intensive scenarios such as RL training, wider bus widths yield significant throughput improvements. Notably, in RL All-Reduce burst tests, the baseline switch suffered severe packet loss during Incast events. In contrast, the optimized design targets Incast ports with increased buffer allocation, maintaining latency within a reasonable range. Overall, compared to fixed-architecture counterparts, the  \sysname switch delivers average latency breakdown of 7.8\%$\thicksim$38.4\% across various tasks. For future work, implementing targeted HLS-based in-switch computing kernels (e.g., All-Reduce aggregation~\cite{iswitch, switchml}) could yield even greater performance gains.

\section{Conclusion}
This paper introduces \sysname, a framework that automates the co-design of custom protocols and FPGA-based switch micro-architectures. We propose a unified DSL-driven workflow that integrates three key innovations: a modular HLS-based switch template allowing flexible composition of protocols and switch architectures; a trace-aware DSE engine for identifying pareto-optimal configurations; and a multi-granularity simulation system that enables hardware-aligned verification within the ns-3 network simulator. Experiments show that this domain-specific adaptation effectively optimizes performance for diverse workloads, achieving resource savings of 55\% in constrained environments and latency reductions of 7.8\%$\thicksim$38.4\% compared to static baselines. These results demonstrate that automated application-specific customization provides a scalable and efficient path for next-generation network infrastructure.
\hfill

\section*{Acknowledgement}
The support of the UK EPSRC (Grant EP/V028251/1, EP/S030069/1, EP/X036006/1), UKRI (Grant 256), KIAT, AMD and Broadcom is gratefully acknowledged.

\bibliographystyle{IEEEtran}
\bibliography{IEEEabrv,references}\balance

\end{document}